\documentclass[a4paper, onecolumn, 11pt, twoside]{article}
\usepackage[dvips]{graphicx}
\usepackage{amsmath}
\usepackage{epsfig}

\newlength{\doublefig} 
\newcommand{\doublefigure}[3][htbp]{
  \begin{figure}[#1]
    \begin{minipage}[t]{\doublefig}
      #2
    \end{minipage}
    \hfill
    \begin{minipage}[t]{\doublefig}
      #3
    \end{minipage}
  \end{figure}
}

\numberwithin{equation}{section}

\date{\today}
\begin{document}

\titlepage

\begin{center}
\vspace{5mm}
\vspace*{2cm}

\begin{flushright}
\begin{minipage}[t]{60mm}
{\bf TSL/ISV-2000-0238 \\
LUNFD6/(NFFL-7191) 2000 \\
December 2000} 
\end{minipage}
\end{flushright}
\end{center}

\vspace*{3.2cm}

\begin{center}
{\Large{Searching for physics beyond the Standard Model in the decay 
$\mathrm{B^+ \rightarrow K^+K^+\pi^-}$ }}\\

\vspace*{2cm}
{J. Damet$^{a}$, P. Eerola$^{b}$, A. Manara$^{c}$, S.E.M. Nooij$^{d,e}$}\\

 \vspace{5mm}
 {\small{
   $^{a}$ Department of Radiation Sciences, Uppsala University, \\
Box 535, S-751 21 Uppsala, Sweden\\
   $^{b}$ Department of Elementary Particle Physics, Lund University, \\
Box 118, S-221 00 Lund, Sweden\\
   $^{c}$ Department of Physics, Indiana University, \\
Bloomington, IN 47405, USA \\
   $^{d}$ Institute for Theoretical Physics, University of Amsterdam,\\
 1018 XE Amsterdam, The Netherlands\\
   $^{e}$ CERN Summer Student\\
 }}
\end{center}

\begin{abstract}
The observation potential of the decay $\mathrm{B^+
\rightarrow K^+K^+\pi^-}$ with the ATLAS detector at LHC is
described in this paper. 
In the Standard Model this decay mode is highly
suppressed, while in models beyond the Standard Model it could
be significantly enhanced. To improve the selection of the 
$\mathrm{K^+K^+\pi^-}$ final state, a charged hadron identification 
using Time-over-Threshold measurements in the ATLAS Transition 
Radiation Tracker was developed and used.
\end{abstract}

\newpage
\section{Introduction}

Many B-meson decays have been considered for observing effects 
originating from physics beyond the Standard Model (SM).  
In general, the following classes of B decays are most 
sensitive to new physics effects:
1) $\Delta b$=1 processes through penguin diagrams,
2) $\Delta b$=2 processes through box diagrams, and 
3) tree-level processes mediated by exchange of a new particle.

Processes such as
${\rm{b \rightarrow s \gamma}}$, belonging to the class 1), 
have been analysed \cite{cleo},
but theoretical uncertainties hamper the observation of new physics 
signatures \cite{hewett}. 
Similar channels such as ${\rm{b \rightarrow s q\bar{q}}}$ \cite{sqq} 
and ${\rm{b \rightarrow s \ell\bar{\ell}}}$ \cite{sll} also suffer from 
large theoretical uncertainties. Some other processes such as 
${\rm{B \rightarrow \tau}}$, representing the class 3),  
have been shown to be rather 
insensitive to a large class of new physics models \cite{btau}.

\begin{figure}[hbt]
\begin{center}
\vspace*{-0.5cm}
\includegraphics[width=140mm]{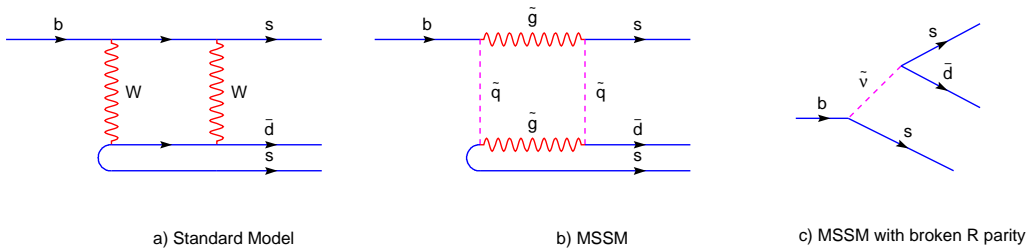}
\vspace*{-0.5cm}
\caption{Feynman diagram contributions to the decay ${\rm{ b \rightarrow
ss \bar{d} }}$ in
various models: a) The Standard Model, b) MSSM, c) MSSM with broken 
R parity.}
\label{feyn}
\end{center}
\end{figure}

Rare decays, representing the class 2), can probe efficiently new physics
effects, since for these decays the SM typically predicts
extremely tiny branching ratios.
The process ${\rm{b \rightarrow s s \bar{d} }}$
is a decay which is strongly suppressed in the SM. This decay can be produced 
in the SM by box diagrams (see Fig.~\ref{feyn}a) 
with an estimated branching ratio at a level lower than $10^{-11}$. 
The Minimal Supersymmetric Standard Model (MSSM) introduces 
squark-gaugino (or higgsino) box diagrams  
(see Fig.~\ref{feyn}b), increasing the estimated branching ratio to  
$10^{-7} - 10^{-8}$ \cite{mssm}. The decay has
also been studied in the context of several Two Higgs Doublet Models (THDM) \cite{thdm}. 
The studies have shown that in these cases, 
the branching ratio could be as high as $10^{-7}$.
Supersymmetry with broken R parity provides another model with a significant 
enhancement of this decay (see Fig.~\ref{feyn}c) \cite{mssm}. These decays are tree-level
processes (class 3) in our classification), and therefore 
the branching ratio could be even significantly higher than those predicted for
the box-diagram processes. 

Three-body decays of a charged B, such as
${\rm{B^\pm \rightarrow K^\pm K^\pm \pi^\mp}}$ (either directly or
through a K$^{*}$-resonance), were suggested as a clear signal for the
process ${\rm{ b \rightarrow ss \bar{d} }}$ in Ref.~\cite{mssm}. 
Recently, an upper limit 
of $8.79\cdot10^{-5}$ was set by the OPAL collaboration 
for the branching ratio BR(${\rm{B^\pm \rightarrow K^\pm K^\pm \pi^\mp}}$)
at $90\%$ confidence level \cite{opal}.

In this paper, the direct decay $\mathrm{B^+ \rightarrow K^+K^+\pi^-}$ was
considered \footnote{Charge conjugated 
states are implicitly included.} in order to test the feasibility
of observing these decays in the ATLAS experiment at the Large Hadron Collider (LHC).
Similar analysis could be performed with a more general class of final
states ${\rm{B^\pm \rightarrow K^\pm K^\pm}}$~+~(no strange), including
also K$^*$-resonances, to increase the statistics. In sections 2-4 the
analysis procedure and simulation results for ATLAS are described. In section 5 
the reach of other experiments is estimated, and section 6 summarises the paper.

\section{Event simulation}
 
The $\mathrm{B^+ \rightarrow K^+K^+\pi^-}$ decay was implemented in
the Monte Carlo program PYTHIA 5.7 \cite{pythia} in
order to generate the signal sample.
In the event generation, b-quark pairs were produced in pp-collisions at 
$\sqrt{s} = 14$ TeV either directly via
the lowest order process, or via gluon splitting or flavour excitation.

Events containing a ${\rm{B^+}}$ meson were selected, and then the
${\rm{B^+}}$ was forced to decay into a $\mathrm{K^+K^+\pi^-}$ final
state. The 
associated ${\rm{\bar{b}}}$ was forced to decay semileptonically into
$\mu$X, in order to satisfy the ATLAS level-1 trigger requirements for B hadrons 
(a muon with a $p_{\rm T}>$~6 GeV and $|\eta| <
2.4$)\footnote{Throughout this paper, the symbol $p_{\rm T}$ is used for 
the transverse momentum with respect to the beam direction, and $\eta$
for the pseudorapidity.}. The muon was not needed in the subsequent
analysis as such.

The ATLAS second level trigger for this hadronic B decay could be envisaged to
be the presence of three charged particles with 
$p_{\rm T}>1.5$~GeV, forming an invariant mass close to the B-meson
mass. The detailed trigger rates have not been studied.

For this feasibility study, a fast simulation program was used instead 
of a full GEANT simulation. The parametrisation was established by studying 
in detail the resolutions of the five helix parameters of the tracks in 
fully-simulated samples, including tails (\cite{fastparam1},\cite{fastparam2}).
The smeared five helix parameters of the track and the corresponding 
covariance matrix were obtained, and a look-up table as a function of 
$p_{\mathrm T}$ and $\eta$ was produced. 
The parametrisation was then applied to the four-momenta of the
generated particles. 
In case of pions, the parametrisation included a dependence on the decay
radius as well, in order to be able to describe
pions coming from the decay of long lifetime particles such as
$\mathrm{K^0_S}$.

\section{Hadron identification}\label{subsec:hadron.identi}

The possibility for separating kaons, protons and pions enhances
the observation potential of many B-hadron final states
in the ATLAS experiment~\cite{bphys}. Monte Carlo 
studies of K/$\pi$ separation using the signal shape information 
from the ATLAS Transition Radiation Tracker (TRT) have been 
previously reported in~\cite{david}.

Recent test-beam data and detailed Monte Carlo simulations 
show that using the time-over-threshold information in the TRT data  
allows for an improved hadron identification, 
assuming that the TRT read-out would also provide the time 
of the trailing edge at low luminosity while preserving 
the output bandwidth requirements.
The time-over-threshold method is described in 
detail in~\cite{nim}.

The time-over-threshold (ToT) for a single straw is 
defined as the width of the signal at the output of the 
low-threshold discriminator in the TRT front-end electronics. 
The ToT provides partial information on the particle energy loss 
in the straw gas, assuming that its dependence on the distance 
of closest approach of the track to the straw anode has been 
taken into account.

The energy loss estimator ($\langle \Delta_{ToT} \rangle$) is built on the 
basis of the individual ToT for all the straw hits on a given particle track, 
according to the procedure described in \cite{nim}. In the TRT, on average, 
35 straws will be crossed by particle tracks with $p_{\mathrm T} > 0.5$~GeV and
$|\eta|< 2.5$. 

\begin{figure}[htb]
\begin{center}
\vspace*{-0.1cm}
\includegraphics[width=85mm]{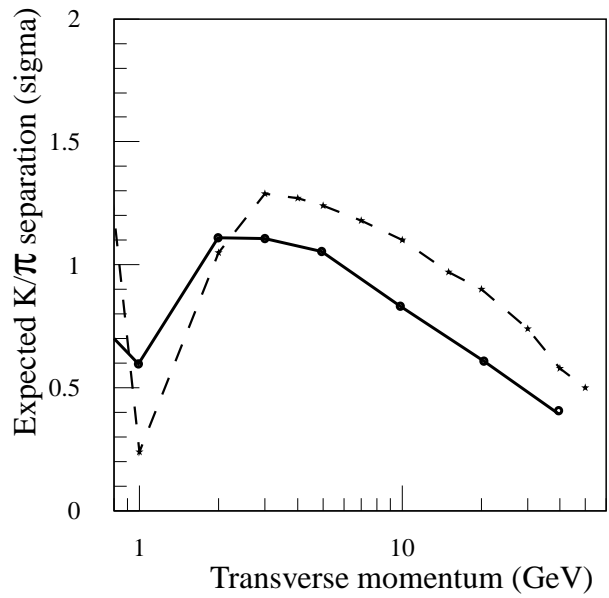}
\vspace*{-0.5cm}
\caption{Expected K/$\pi$ separation in units of standard deviation
in the ATLAS TRT as a function of transverse 
momentum (no pile-up effects included). The separation is shown as an average over 
the full rapidity coverage (solid line) and at $|\eta| = 0.3 $ (dotted line).}
\label{fig:kapi4}
\end{center}
\end{figure}

The expected K/$\pi$ separation as a function of 
the transverse momentum 
is shown in Fig.~\ref{fig:kapi4} in units of standard deviation. 
Without including any pile-up effects, 
the K/$\pi$ separation is predicted to be above one standard deviation
for transverse momenta between 2 and 5\,GeV, 
averaged over the full rapidity coverage (solid line), 
and above one standard deviation over a broader $p_{\mathrm T}$-range
between 2 and 15\,GeV at $|\eta| = 0.3 $ (dotted line).

In order to study physics processes in the ATLAS experiment, 
the mean and the sigma of the $\langle \Delta_{ToT} \rangle$ distributions, 
well described by gaussians, were parameterised for pions, kaons and protons 
as a function of $p_\mathrm{T}$ and $\eta$ over the full acceptance
region of the ATLAS TRT \cite{atlasnote}.

\clearpage

\section{Analysis}
\subsection{Event selection}\label{subsec:signal.selection}

The event selection cuts are summarised in Table~\ref{tab:cuts}.
The cuts on the transverse momentum of the particles and the loose cut
on the B-candidate mass emulated the suggested second level trigger requirements.
The other selection criteria were based on the quality of the B-vertex fit,
on the long lifetime of the B meson (the reconstructed B vertex was
required to be separated from the primary vertex with at least
100~$\mu$m in the transverse plane), on the rejection of events in which two of the particle 
pairs formed masses close to light resonance masses, and on the 
probability that the three particles formed a KK$\pi$ combination.
The last selection criterium is explained in more detail in
Section \ref{subsec:tot}. The resolution of the decay length in the transverse 
plane was 72\,$\mu$m.

\begin{table}[hbt]
\begin{center}
\begin{tabular}{|l|c|c|c|}     \hline
No.&  Selection               & Signal  &  Background  \\   
   &                          & efficiency &  efficiency \\  \hline 
1. &  $p_T$(tracks) $>$ 1.5\,GeV     &      50.9\%       & 68.8\% \\ \hline
2. &  4~GeV $< M$(B) $<$6~GeV        &     97.7\%       & 53.1\% \\  \hline
3. &  $\chi^2({\rm{triplet~vertex~fit}}) < 2$&   83.6\% & 68.4\% \\  \hline
4. &  $p{\mathrm{_T}({\rm{B}}) > 10}$ GeV &      80.1\% & 22.6\% \\  \hline 
5. &  Vertex detachment $>$  0.1mm   &      58.3\% &  0.3\% \\  \hline
6. &  $\mathcal{P}(dE/dx) > 0.1$     &      87.8\%  & 76.3\% \\  \hline
7. &  $m_{13}^2$ and $m_{23}^2$ $>$ 2.5 GeV$^2$ &  74.4\% & 35.9\% \\  \hline
8. & ${\rm{5.16~GeV < M(B) < 5.45~GeV}}$ &  91.2\% &  5.8\% \\  \hline \hline
   &  Overall                        &       11.6\% & $2.8\cdot 10^{-4}$ \%   \\  \hline

\end{tabular}
\end{center}
\caption{The signal and background efficiency of each selection cut. 
The cuts 1-8 were applied in sequence, and the efficiency of cut $N$ is given relative
to the remaining sample ($N-1$ cuts applied).}
\label{tab:cuts}
\end{table}

The overall signal efficiency was found to be 11.6\%, while the background
efficiency was $2.8\cdot 10^{-4}$~\%. The study of the background rejection
was limited by the statistics of the simulated background sample (one
million inclusive $\mathrm{b\bar{b} \rightarrow \mu 6 X}$ events, where $\mu$6
denotes the level-1 trigger requirements for the muon). The generated background 
consisted of the default decay channels in PYTHIA 5.7.

\subsection{Use of the $dE/dx$ information}\label{subsec:tot}

The selection criteria discussed in Section \ref{subsec:signal.selection}
reduce the background by six orders of magnitude,
while preserving about $12\%$ of the signal,
as it is shown in Table~\ref{tab:cuts}.
In this section, the use of the $dE/dx$ information is explained in more
detail.

For a charged particle track with a given $p_\mathrm{T}$ and $\eta$,
the \emph{$dE/dx_{actual}$}($p_\mathrm{T},\eta$)
was simulated using the sigma and the mean provided by the
parametrisation (see Section \ref{subsec:hadron.identi}):

\begin{eqnarray*}
  \frac{dE}{dx}\Big|_{actual}(p_\mathrm{T},\eta) = \text{mean}(\langle \Delta_{ToT} \rangle)
(p_\mathrm{T},\eta)
 + \text{{\sc rnd}} * \sigma(\langle \Delta_{ToT} \rangle)(p_\mathrm{T},\eta),
\end{eqnarray*}
where RND is a gaussian-distributed pseudo-random number.
  
For any given triplet of particles, two positively and one negatively charged ---
candidates for the decay products of the $B^+$ --- the
$\chi^2$ distribution was constructed according to :
\begin{eqnarray*}
\chi^{2} = \sum_{i=1}^{3}\Bigg[\frac{\frac{dE}{dx}\big|_{exp}-\frac{dE_i}{dx}\big|_{act}}{\sigma_i}\Bigg]^{2}
\end{eqnarray*} 
where the index $i$ labels the individual particles
in the triplet and $\frac{dE}{dx}|_{exp}$ is the mean value of the $dE/dx$
distribution for pions (if the particle had a negative charge)
or for kaons (if the particle had a positive charge).
The obtained $\chi^2$ probability distribution 
for three degrees of freedom is shown in Fig.~\ref{fig:chisq2}.
The background misidentification probability as a function
of the signal efficiency is shown in Fig.~\ref{fig:dedx_eff}, when the cut
on the $dE/dx$ $\chi^2$ probability was varied. 

\doublefigure[htbp]{
  \begin{center}
\epsfig{file=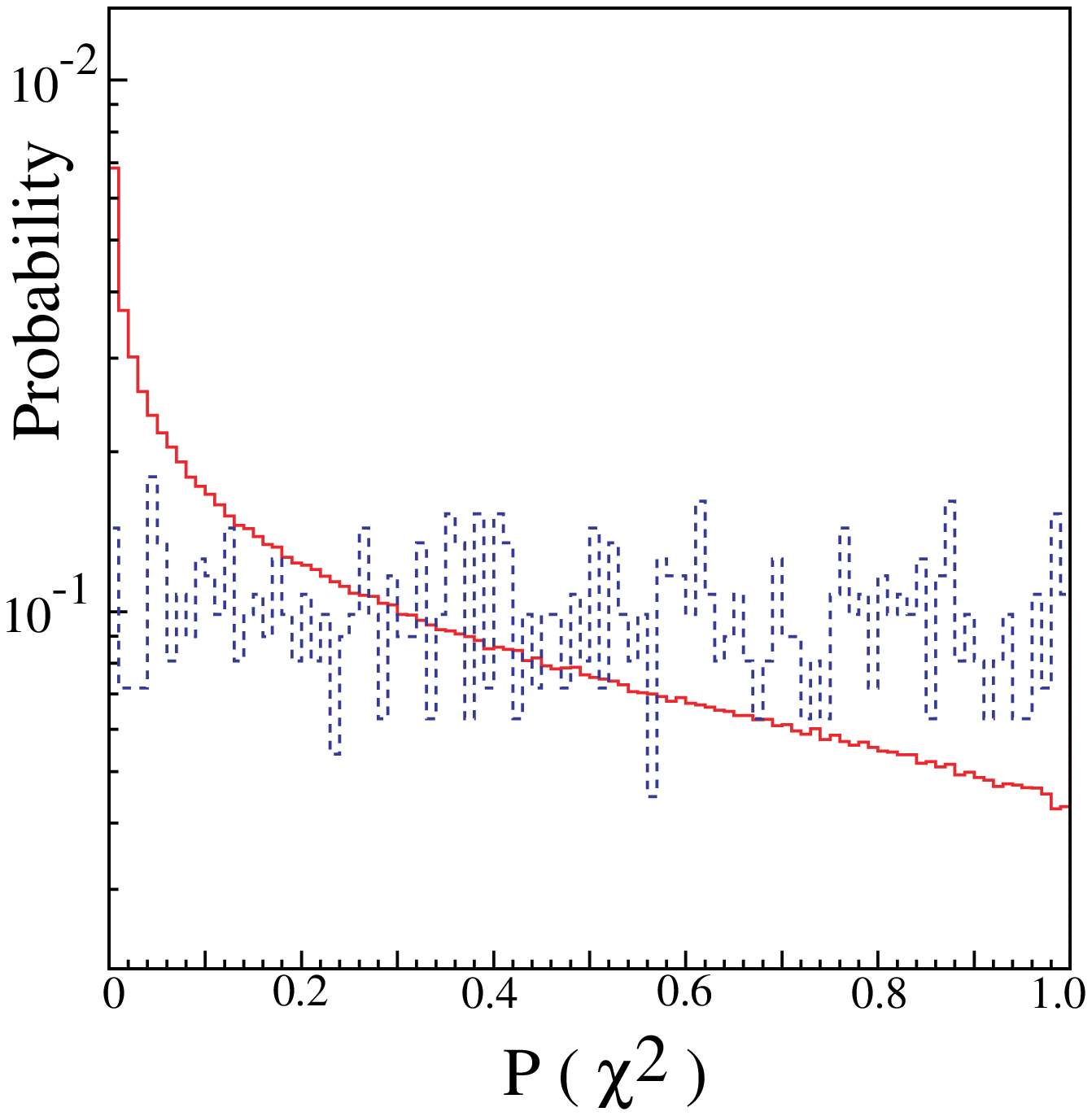,width=\doublefig}
  \caption{The $\chi^2$ probability distribution for three degrees of freedom for the
signal (dotted line) and the background (solid line), after applying 
the proposed second level trigger requirements (see Sect. 2). Both distributions were normalised 
to unity.}
  \label{fig:chisq2}
  \end{center}
}{
  \begin{center}
   \epsfig{file=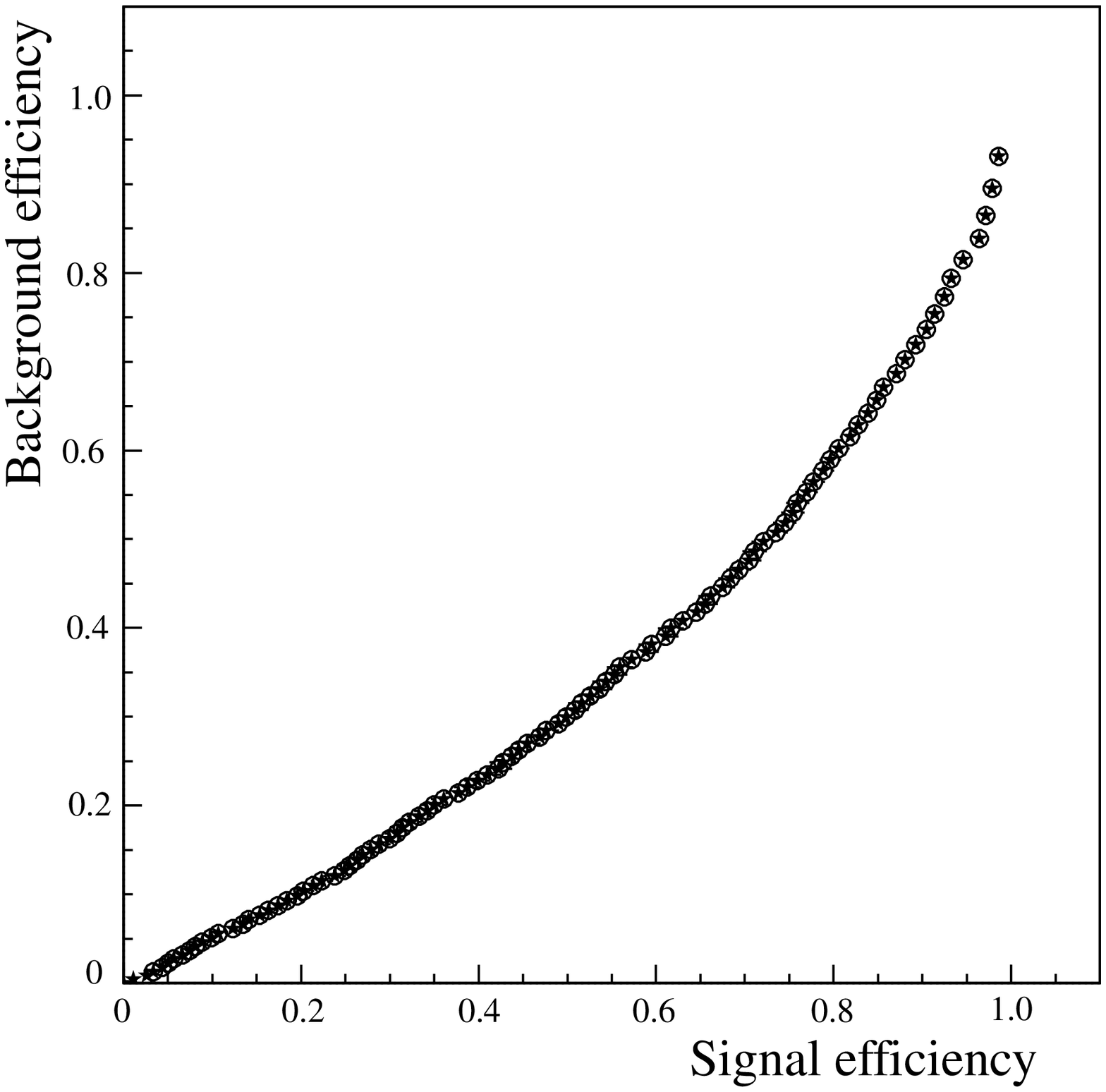,width=\doublefig}
   \caption{Background misidentification probability as a function of the
signal efficiency, when the cut
on the $dE/dx$ $\chi^2$ probability was varied. 
The proposed second level trigger requirements were applied to the event sample first (see Sect. 2).}
    \label{fig:dedx_eff}
  \end{center}
}

\subsection{Results}

The number of signal events, passing the ATLAS level-1 trigger, was estimated as:

\begin{eqnarray*}
N\mathrm{_{signal}^{prod} = \sigma(pp \rightarrow b\bar{b} 
\rightarrow \mu6X) \cdot
\mathcal{B}r (b \rightarrow B^+) \cdot \mathcal{B}r 
(B^+ \rightarrow K^+K^+\pi^-) \cdot \int \mathcal{L}dt},
\end{eqnarray*}
and the number of observed events as:
\begin{eqnarray*}
N\mathrm{_{signal}^{obs}} =  N\mathrm{_{signal}^{prod} 
\cdot \epsilon_{rec} \cdot \epsilon_{id} },
\end{eqnarray*}
where the cross-section after the level-1 trigger is 
$\sigma({\mathrm{pp \rightarrow b\bar{b} \rightarrow \mu 6 X) = 2.3 \mu b}}$,
$ \mathcal{B}r {\mathrm{ (b \rightarrow B^+) = 39.7\%}}$,
the integrated luminosity is $ \int \mathcal{L}dt = {\mathrm{30 fb^{-1} }}$ 
(corresponding to three years of LHC data-taking at the initial low luminosity),
the signal reconstruction efficiency is $ \epsilon_{\rm rec} = 11.6 \%$,
the muon efficiency is $ \epsilon_{\rm id}(\mu6) = 0.85$, and the pion and kaon 
reconstruction efficiencies are
$ \epsilon_{\rm id}(\pi,\mathrm{K}) = 0.90$.

The number of background events was estimated as:

\begin{eqnarray*}
N\mathrm{_{bg}^{prod} = \sigma(pp \rightarrow b\bar{b} 
\rightarrow \mu6X) \cdot \int \mathcal{L}dt }.
\end{eqnarray*}
The number of produced background events was thus
$N\rm{_{bg}^{prod} = 6.9 \cdot 10^{10}}$ for 30 fb$^{-1}$. 
Using $ \epsilon_{\rm rec} = 2.8\ 10^{-4} \% $ and 
$ \epsilon_{\rm id}(\mu6) = 0.85$,
$ \epsilon_{\rm id}(\pi,\mathrm{K}) = 0.90 $, the number of 
observed background events was $N\rm{_{bg}^{obs} = 1.2 \cdot 10^{5} }$.

An upper limit of
$$ \mathcal{B}r {\mathrm{ (B^+\rightarrow K^+K^+\pi^-) = 3.4 \cdot 10^{-7} }}$$
can be established at a 95\% CL after three years of data-taking at the low
luminosity.
If one requires a signal significance of five standard deviations, 
a signal with a branching ratio of:
$$ \mathcal{B}r {\mathrm{ (B^+\rightarrow K^+K^+\pi^-) = 8.8 \cdot 10^{-7} }}$$
can be observed with the same statistics.

\subsection{Limits on the R-parity violating couplings}

The $\mathrm{b \rightarrow ss\bar{d}}$ decay rate induced by the R-parity violating couplings 
was estimated in Ref.~\cite{mssm} to be :

\begin{eqnarray*}
\mathrm{\Gamma(b \rightarrow ss\bar{d}) = \frac{m_b^5 f_{QCD}^2}{512 (2\pi)^3 m_{\bar{\nu}}^2} 
\left( 
\left| \sum_{i=1}^3 \lambda_{i32}^\prime \lambda_{i21}^{\prime *} \right| ^2
+
\left| \sum_{i=1}^3 \lambda_{i12}^\prime \lambda_{i23}^{\prime *} \right| ^2
\right)}, \label{eq:lambda}
\end{eqnarray*}

where $m_{\mathrm b}$ is the b-quark mass, 
$f_{\mathrm{QCD}}=(\alpha_s({\mathrm b})/\alpha_s(m_{\bar{\nu}}))^{24/23}$, 
$m_{\bar{\nu}}$ is the sneutrino mass and $\lambda^\prime$ are dimensionless couplings.

It was estimated in Ref.~\cite{mssm} that a 
fourth (or less) of the $\mathrm{b \rightarrow ss\bar{d}}$ transitions leads 
to ${\rm{B^\pm \rightarrow K^\pm K^\pm}}$ + (no strange) decay channels. 
Final states including ${\rm{K^\pm K^\pm}}$ + (no strange) can be produced 
via one or two excited 
kaons with respective proportions ${\rm{K^*K^* \geq K^*K \geq KK}}$. It
was assumed, pessimistically, that direct ${\rm{K^\pm K^\pm}}$ decay represents 
a third of the total 
${\rm{K^\pm K^\pm}}$ + (no strange) decays, 
thus 1/12 of the $\mathrm{b \rightarrow ss\bar{d}}$ 
transitions. Using this estimation with the 95 \% CL ATLAS bound and, as in Ref.~\cite{mssm}, 
$m_{\mathrm b}$ = 4.5 GeV, $f_{\mathrm{QCD}}$ = 2, 
$m_{\bar{\nu}}$ = 100 GeV, 
${\rm{\tau_{B^+}\ =\ 1.65\ ps}}$, a limit on the couplings will be:

$$ \sqrt{ \left|\sum_{i=1}^3 \lambda_{i32}^\prime \lambda_{i21}^{\prime *} \right| ^2
+ \left| \sum_{i=1}^3 \lambda_{i12}^\prime \lambda_{i23}^{\prime *} \right| ^2} 
< 5.3 \cdot 10^{-5}.$$

This limit should be considered as a rough estimate, obtained using very pessimistic
assumptions on the relative decay probabilities in order to maximize the relative
fraction of the decay mode $\mathrm{B^+\rightarrow K^+K^+\pi^-}$.  The limit is
nevertheless an order of magnitude better than the corresponding OPAL limit in
Ref.~\cite{opal}. Complementary
measurements come from neutrino data \cite{sklim} which give values for individual
$\lambda^\prime$ couplings for different neutrino mass scenarios. 

\section{Comparison to the other experiments}

If the branching ratio of the decay $\mathrm{B^+\rightarrow K^+K^+\pi^-}$ is in the range
of the MSSM or THDM predictions ($\mathcal{O}(10^{-7})$), 
the event yield of the PEP-II and KEKB B-factories is only a few events 
with an integrated luminosity of 30 fb$^{-1}$ to 100 fb$^{-1}$.
Therefore these investigations would not seem feasible 
for BaBar and Belle in the R-parity conserving scenarios.

The hadron colliders Tevatron and LHC have much larger ${\rm{b\bar{b}}}$ 
cross sections, which opens up the opportunity to study 
${\rm{b \rightarrow ss\bar{d}}}$ transitions.
For CDF it was assumed that the trigger efficiency is a ten times better than 
the ATLAS trigger efficiency, due to the possibility of triggering on purely hadronic final 
states. It was assumed that CDF has the 
same signal reconstruction efficiency as ATLAS. The K/$\pi$ separation capability is similar
in the two experiments for transverse momenta above 1.3 GeV. ATLAS has a larger 
pseudorapidity acceptance than CDF, but on the other hand the final states are more central
with less initial state gluon radiation at the smaller center-of-mass energy. 
The estimated overall efficiency 
was consistent with the results on experimentally similar decays \cite{cdfb99}.
Using these assumptions, one can estimate that CDF could set an upper limit of 
$7.9 \cdot 10^{-7}$ at 95\% CL with 2 fb$^{-1}$. A signal significance of five standard
deviations could be achieved if the branching ratio were $2.0 \cdot 10^{-6}$. 

The LHCb experiment has the advantage of being able to trigger on purely hadronic final
states, and having a superior K/$\pi$ separation 
thanks to its RICH detectors. Based on the results presented in Table 15.11 in 
Ref.~\cite{lhcb}, it was estimated 
that LHCb could observe a five-standard-deviation signal if the branching
ratio were
$1.0 \cdot 10^{-7}$, given the statistics of three years' running 
at the nominal
LHCb luminosity of $2 \cdot 10^{-32}$~cm$^{-2}$~s$^{-1}$ .
The 95\% CL upper limit for the branching ratio would be $4.0 \cdot 10^{-8}$.

These results should be taken as crude estimations, which were based on the publicly 
available information on the detector and accelerator performance.

\section{Summary and outlook}

A feasibility study of reconstructing 
the decay ${\mathrm{B^+\rightarrow K^+K^+\pi^- }}$ in the ATLAS experiment at LHC has
been presented. The obtained 95\% CL upper limit of
$$ \mathcal{B}r {\mathrm{ (B^+\rightarrow K^+K^+\pi^-) = 3.4 \cdot 10^{-7} }}$$
approaches the range of branching fractions predicted by
MSSM or THDM scenarios. In R-parity violating models, branching ratios could be as
large as $10^{-4}$. ATLAS could thus
contribute in the measurements of some of the
R-parity violating couplings. Given the upper limit above, the following limit can be
set on the couplings:
$$ \sqrt{ \left|\sum_{i=1}^3 \lambda_{i32}^\prime \lambda_{i21}^{\prime *} \right| ^2
+ \left| \sum_{i=1}^3 \lambda_{i12}^\prime \lambda_{i23}^{\prime *} \right| ^2} 
< 5.3 \cdot 10^{-5}.$$

The presented analysis of the $\mathrm{B^+ \rightarrow K^+K^+\pi^-}$ decay shows that
the ATLAS experiment will be able to set a new upper limit on the branching ratio, 
which will be more than two orders of magnitude lower than the present estimate. 
The limit on the relation
that constrains the $\lambda^\prime$ couplings of the MSSM with R-parity violation
will also be improved by an order of magnitude. This analysis considered only 
the direct decay ${\mathrm{B^+\rightarrow K^+K^+\pi^- }}$, but ATLAS should be able 
to increase the statistics by
searching for final states with ${\rm{K^*K^*}}$ and ${\rm{K^*K}}$.

Combining all the results, the LHC experiments will contribute significantly to the search 
and measurements of physics beyond the Standard Model using the 
$\mathrm{B^+ \rightarrow K^+K^+\pi^-}$ decay channel. 

\section{Acknowledgements}
This work has been performed within the ATLAS Collaboration, and we
thank collaboration members for helpful discussions.
We have made use of the physics analysis framework and tools which are
the result of collaboration-wide efforts.
The authors would like to thank J. Bijnens and K. Huitu for 
helpful discussions on the theoretical aspects, and
D. Rousseau for fruitful discussions on energy loss techniques for the TRT.
A. Manara would also like to thank H. Ogren and F. Luehring for 
several useful discussions.


\newpage
\begin{thebibliography}{40}
\bibitem {cleo} CLEO Collaboration, R. Ammar {\it{et al.}}, Phys. Rev.
Lett. 71 (1993) 674;\\
CLEO Collaboration, M.S. Alam {\it{et al.}}, Phys. Rev. Lett. 74 (1995)
2885.
\bibitem {hewett} J.L. Hewett, Phys. Rev. Lett. 70 (1993) 1045.
\bibitem {sqq} L.L. Chau, H.Y. Cheng, W.K. Sze, H. Yao and B. Tseng,
Phys. Rev. D 43 (1991) 2176;\\
               C.-D. L\"u and D.-X. Zhang, Phys. Lett. B 400 (1997) 188.
\bibitem {sll}  C.-D. L\"u and D.-X. Zhang,  Phys. Lett. B 397
 (1997) 279;\\
G. Buchalla and G. Isidori, Nucl. Phys. B 525 (1998) 333.
\bibitem {btau} D. Guetta and E. Nardi, Phys. Rev. D 58 (1998) 012001.
\bibitem {mssm} K. Huitu, C.-D. L\"u, P. Singer and D.-X. Zhang, Phys.
Rev. Lett. 81 (1998) 4313.
\bibitem {thdm} K. Huitu, C.-D. L\"u, P. Singer and D.-X. Zhang, Phys.
Lett. B 445 (1999) 394.
\bibitem {opal} OPAL Collaboration, G. Abbiendi {\it{et al.}},
Phys. Lett. B 476 (2000) 233. 
\bibitem {pythia} T. Sj\"ostrand, Computer Physics Commun. 82 (1994) 74.
\bibitem {fastparam1} E. J. Buis, R. Dankers, S. Haywood and A. Reichold,
ATLAS Internal Note ATL--INDET--97--195 (1997). 
\bibitem {fastparam2} E. J. Buis {\it{et al.}}, ATLAS Internal
Note ATL--INDET--98--215 (1998). 
\bibitem{bphys} ATLAS Collaboration, 
ATLAS Detector and Physics Performance Technical Design Report Vol II,
CERN/LHCC/99--15 (1999).
\bibitem{david} ATLAS Collaboration, 
ATLAS Detector and Physics Performance Technical Design Report Vol I, 
CERN/LHCC/99--14 (1999).
\bibitem{nim} T. Akesson {\it{et al.}},
`Particle Identification using the Time-over-Threshold Method 
in the ATLAS Transition Radiation Tracker', ATLAS Internal Note
ATL--INDET--2000--021 (2000), submitted to Nucl. Instr. 
and Methods A.
\bibitem{atlasnote} J. Damet, P. Eerola, A. Manara and S.E.M. Nooij,
ATLAS Internal Note ATL--PHYS--2000--027 (2000).
\bibitem{sklim} G. Bhattacharyya, Phys. Rev. D 59 (1999) 091701.
\bibitem{cdfb99} V. Papadimitriou, Nucl. Instrum. and Methods A 446 (2000) 143.
\bibitem{lhcb} LHCb Collaboration, LHCb Technical Proposal, CERN/LHCC/98--1, pp. 155--157.
\end{thebibliography}
\end{document}